\documentclass{aastex62}

\usepackage{amsmath}

\graphicspath{{./}{figures/}}

\received{--}
\revised{--}
\accepted{--}
\submitjournal{ApJ}

\shorttitle{Global NSSL modeling}
\shortauthors{Stejko et al.}

\begin{document}

\title{3D MHD Modeling of the Impact of Subsurface Stratification on the Solar Dynamo}


\correspondingauthor{Andrey Stejko}
\email{ams226@njit.edu}

\author{Andrey M. Stejko}
\affiliation{New Jersey Institute of Technology\\
 323 Dr Martin Luther King Jr Blvd.\\
 Newark, NJ 07012, USA}

\author{Gustavo Guerrero}
\affiliation{Universidade Federal de Minas Gerais\\
Av. Antonio Carlos, 6627\\
Belo Horizonte, MG, 31270-901, Brazil}

\author{Alexander G. Kosovichev}
\affiliation{New Jersey Institute of Technology\\
 323 Dr Martin Luther King Jr Blvd.\\
 Newark, NJ 07012, USA}

\author{Piotr K. Smolarkiewicz}
\affiliation{European Center For Medium Range Weather Forecasts\\
 Shinfield Rd\\
 Reading RG2 9AX, UK}

\begin{abstract}

\indent Various models of solar subsurface stratification are tested in the global EULAG-MHD solver to simulate diverse regimes of near-surface convective transport. Sub- and superadiabacity are altered at the surface of the model ($ r > 0.95~R_{\odot}$) to either suppress or enhance convective flow speeds in an effort to investigate the impact of the near-surface layer on global dynamics. A major consequence of increasing surface convection rates appears to be a significant alteration of the distribution of angular momentum, especially below the tachocline where the rotational frequency predominantly increases at higher latitudes. These hydrodynamic changes correspond to large shifts in the development of the current helicity in this stable layer ($r<0.72R_{\odot}$), significantly altering its impact on the generation of poloidal and toroidal fields at the tachocline and below, acting as a major contributor towards transitions in the dynamo cycle. The enhanced near-surface flow speed manifests in a global shift of the toroidal field ($B_{\phi}$) in the butterfly diagram - from a North-South symmetric pattern to a staggered anti-symmetric emergence.

\end{abstract}

\keywords{Solar Physics, Computational Fluid Dynamics, 3D Solar Modeling}


\section{Introduction} \label{sec:intro}

\indent Dynamo theory has attempted to model observations of the 22-year solar magnetic cycle. By employing large-scale motions like differential rotation and/or meridional flows, together with the contribution of turbulent cyclonic motions and currents, the so called $\alpha$-effect, dynamo models aim to simulate action of the large-scale solar magnetic field \citep{2010LRSP....7....3C}. Polarity reversals every 11-years, equatorward migration of sunspots during cycles as well as the hemispheric polarity rule, with an anti-symmetric magnetic field across the equator, are issues that still have not been adequately explained. Between mean-field dynamo-wave \citep{1955ApJ...122..293P,1978mfge.book.....M,2005PhR...417....1B,2019arXiv190804525P} and flux-transport \citep{1961ApJ...133..572B,1969ApJ...156....1L} solar dynamo models there is still no consensus about the locations and mechanisms responsible for these observations.\\
\indent In mean-field dynamo theory, the generation of the toroidal magnetic field is presumed to be dominant in the high shear region of the tachocline, but the location of the $\alpha$-effect is particularly uncertain. \cite{1955ApJ...122..293P} and \cite{1966ZNatA..21..369S} theorized the contribution of the
turbulent kinetic helicty, or $\alpha$-effect, while \citet{1961ApJ...133..572B} suggested the formation of a near-surface poloidal field due to the diffusive decay of active regions. In this scenario the connection with the toroidal field at the tachocline would occur via meridional circulation. This approach found relative success in modeling some of the main characteristics of the solar cycle for profiles of differential rotation, meridional circulation and the kinetic $\alpha$-effect \citep{[e.g.][]1999ApJ...518..508D, 2002Sci...296.1671N}. These simulations were performed in one hemisphere and anti-symmetry of the toroidal field across the equator was not explored. This parity rule resulted in an issue for flux-transport dynamo models when both hemispheres were considered. Several alternate theories have been since proposed to address the issue -- among these are three hypotheses: placing the $\alpha$-effect near the tachocline \citep{2001ApJ...559..428D, 2002A&A...390..673B}, considering a magnetic diffusivity for the poloidal field two orders of magnitude larger than that of the toroidal field \citep{2004A&A...427.1019C}, and including the contribution of turbulent pumping \citep{2008A&A...485..267G}. The first hypothesis has some caveats since the strong toroidal field in the tachocline would normally quench kinetic turbulent action. Nevertheless, the development of buoyant or magneto-shear instabilities in this region may circumvent this problem, providing a different source for the $\alpha$-effect \citep{2001ApJ...559..428D}. By using the test-field method to explore the non-linear behavior of the turbulent coefficients in simplified flows, \cite{2014ApJ...795...16K} found that the quenching of turbulent diffusivity is isotropic, i.e., the same for all field components. Even though the conditions in the test-field models differ from the solar interior, the second hypothesis seems unrealistic in view of these results. The third hypothesis is still feasible, however, it is hard to predict the amplitudes and profiles of the turbulent transport coefficients. 3D global non-linear solar modeling has attempted to naturally answer some of these questions by incorporating realistic conditions of observable solar parameters in simulations of internal solar dynamics \citep{2004ApJ...614.1073B, 2013ApJ...777L..29C}. Even though there has been enormous progress since the seminal work of \cite{1981ApJS...46..211G}, models are not yet able to successfully reproduce many solar cycle properties; migration patterns of the surface field as well as a hemispheric parity of the solutions has yet to be naturally simulated in a satisfactory way.\\
\indent More recent non-linear global simulations of particular relevance to the current paper have studied the effect of the tachocline on the long-term evolution of the magnetic solar cycle \citep{2016ApJ...819..104G}. Their model included a sub-adiabatic region under the convection zone to mimic the transition from the radiative interior. The addition of this layer sees a dramatic increase in the period of the dynamo cycle, pushing it from a time-frame of years \citep[see also][]{2015ApJ...809..149A, 2018A&A...616A..72W} to that of decades. The strong oscillatory magnetic field ($\sim$ 1 Tesla) forms and is stored in the tachocline and below. These models also form the beginnings of a near-surface shear layer (NSSL) in higher Rossby number regimes where buoyancy overcomes the Coriolis force. High levels of shear are instrumental in generating large-scale toroidal fields, seen dominating the surface and the base of the tachocline in previous models \citep{2016ApJ...819..104G}. These simulations exhibit solar cycles that strongly rely on varying regions of shear as well as turbulent helicities to generate large-scale poloidal fields. The model explored in this paper uses a spherically isomorphic forcing function along with a constant rotation rate to set the conditions for generating solar-like differential rotation and convective flow speeds compatible with the solar interior -- a set of variables that organically distribute variations of turbulence and shear throughout the domain of the model. While previous works \citep{2013ApJ...779..176G,2016ApJ...819..104G, 2019ApJ...880....6G} have explored altering rotation rates, this paper examines effects of changing convective flow patterns by modifying the near-surface stratification of the model -- a region whose subsurface rotational shear, observed by results of helioseismology \citep{kosovichev97,1998ApJ...505..390S}, plays a key role in shaping the emergence of the solar magnetic field in the form of the butterfly diagram \citep{2005ApJ...625..539B, 2011ApJ...727L..45P}. In this work we introduce changes to the subsurface boundary layer and explore its impact on internal dynamics. Since our changes modify the parity of the solution, we explore the conditions that lead to that change.\\
\indent This paper is organized as follows. In \S\ref{sec:model} we describe the numerical model and the internal stratification imposed in our simulations. In \S\ref{sec:results} we present our results and analysis and state our concluding remarks in \S\ref{sec:conclusions}. 

\section{Model Description} \label{sec:model}
\subsection{Model Background}

EULAG-MHD \citep{smolar13} is an extension of the hydrodynamic model EULAG predominantly used in atmospheric and climate research \citep{prusa08}. It is a versatile numerical solver, well adapted to simulating high-Reynolds number anelastic flows found in the majority of the solar interior. EULAG is powered by a nonoscillatory forward-in-time MPDATA method (multidimensional positive definite advection transport algorithm; see \citet{smolar06} for an overview)---a nonlinear second-order-accurate iterative implementation of the elementary first-order-accurate flux-form upwind scheme. The leading truncation terms of MPDATA were shown to act as effective subgrid-scale (SGS) turbulence models \citep{smpr02}. Usually, solar modeling has relied on explicit SGS turbulence models to simulate the transfer and dissipation of energy below the inertial subrange by means of large eddy simulation (LES) \citep{lilly66}. A newer class of LES methods has also been implemented with the dynamic Smagorinsky model \citep{germano91, 2013ApJ...762...73N,2015arXiv150707999W}, increasing the accuracy of turbulent dissipation calculations. EULAG offers an alternative to these classic methods by exploiting the so-called ``implicit large eddy simulation'' (ILES) approach \citep{iles07}, which achieves the same goal while obviating the need for evaluating higher-order differential operators. Other nonoscillatory advection schemes possess the ILES property, and generally have proven to be effective through a large range of scales and physical scenarios, from laboratory to stellar \citep{iles07}. In contrast to classic LES allowing to prescribe the filter length scale, the MPDATA-based ILES is parameter free and adaptive to the solution regularity \citep{margolin06}, while confined to the scales of grid resolution \citep{margolin02,Domar03,kuhnlein19}. The latter minimises viscous dissipation \citep{Domar03,Strugarek16}, and can be helpful to capture bifurcations sensitive to small perturbations in simulation of flows with multiple dynamic equilibria; cf. \cite{Kumar15}. The MPDATA based ILES has been extended in EULAG to global solar convection and compared with the explicitly filtered spectral methods \citep{elliott02} and tested for various solar regimes, reproducing hydrodynamic effects observed on the Sun, such as breaking the Taylor-Proudman balance as well as differential rotation \citep{2013ApJ...779..176G}. This model has been augmented for simulating the global solar magnetic dynamo in the EULAG-MHD version \citep{smolar13} adding an ideal MHD component, and used to study global magnetic dynamics \citep{2010ApJ...715L.133G,Racine11}.\\
\indent EULAG-MHD is described in full detail by \cite{smolar13}, and its conservative properties are thoroughly discussed by \cite{2017ApJ...841...65C}. Here we only briefly summarise its mathematical formulation and specific setup, following \cite{2016ApJ...819..104G}. The governing equations (\ref{eq:gov1})-(\ref{eq:gov4}) assume the anelastic formulation of \citet{LH82}. They are solved in a global spherical shell $0 \le \phi \le 2\pi$, $0 \le \theta \le \pi$, from the radiative interior at $0.61 R_{\odot}$ to the surface at $0.97 R_{\odot}$. The grid resolution $128 \times 64\times 64$ in $\phi$, $\theta$ and $r$, respectively, corresponds to that of previous long-term EULAG simulations of stellar climates \citep{2013ApJ...779..176G,2016ApJ...819..104G}. The governing equations are

\vskip-0.6cm

\begin{gather}
  \nabla \cdot (\rho_{s} \mathbf{u}) = 0\;, \label{eq:gov1}\\
  \dfrac{D \mathbf{u}}{Dt} + 2\mathbf{\Omega}\times\mathbf{u} = - 
\nabla \left(\dfrac{p'}{\rho_{s}}\right) 
+ \mathbf{g}\left(\dfrac{\Theta'}{\Theta_{s}}\right) 
+ \dfrac{1}{\mu_{0}\rho_{s}}(\mathbf{B} \cdot \nabla)\mathbf{B}\;, \label{eq:gov2}\\
  \dfrac{D\Theta'}{Dt} = -\mathbf{u} \cdot \nabla \Theta_{e} 
- \dfrac{\Theta'}{\tau}\;, \label{eq:gov3}\\
  \dfrac{D\mathbf{B}}{Dt}= (\mathbf{B} \cdot \nabla)\mathbf{u}
 - \mathbf{B}(\nabla \cdot \mathbf{u})\;. \label{eq:gov4}
\end{gather}

\indent In equations (\ref{eq:gov1})-(\ref{eq:gov4}), $D/Dt = \partial/\partial t + {\bf u}\cdot\nabla$, where $\mathbf{u}$ denotes the flow velocity, and $\mathbf{\Omega}=\Omega_0(\cos\theta,-\sin\theta,0)$ is the angular velocity of the rotating reference frame ($\Omega_0= 2.6 \times 10^{-6}$). Furthermore, $p$ and $\Theta$ mark pressure and potential temperature, with the latter tantamount to the specific entropy of an ideal gas via $ds=c_p d\ln\Theta$ where $c_p$ is a specific heat at constant pressure. Primes refer to perturbations about a static ambient state assumed to satisfy the generic Lipps-Hemler equations, an asymptotic expansion about the hydrostatic isentropic background state (denoted by the subscript ``s'') under gravitational acceleration $\mathbf{g} \propto r^{-2}$; see \cite{2017ApJ...841...65C} for a substantive discussion. The evolution of the magnetic field, $\mathbf{B}$, is governed by the induction equation (\ref{eq:gov4}) in the classic ideal MHD limit, where the momentum equation (\ref{eq:gov2}) includes the magnetic tension (${\mu_{0}^{-1}}(\mathbf{B} \cdot \nabla)\mathbf{B}$) portion of the Lorentz force, with the magnetic pressure subsumed in $p'$ \citep{smolar13}. The explicit viscous tensor and the magnetic dissipation term are disregarded in favor of the ILES dissipative action of MPDATA \citep{2010ApJ...715L.133G}.\\
\indent With this mathematical design, convection is primarily driven by the advection of a non-isentropic $\Theta_{e}$; i.e., the first term on the rhs of equation (\ref{eq:gov3}). This ambient state can be used to create super-adiabatic entropy gradients that simulate heat flux from the bottom boundary. If left to evolve naturally, convection will tend to balance out any entropy gradient in the ambient state. To offset fluid homogenization, the Newtonian cooling term (the second term on the rhs of equation (\ref{eq:gov3})) relaxes potential temperature perturbations to zero, driving the system towards a balanced statistical equilibrium. This can be viewed as a scale independent parameterisation of the Reynolds heat flux \citep{2017ApJ...841...65C}. In these simulations, the timescale of the relaxation is set at $\tau = 1.036\times10^{8}\text{ s} \approx 3.3 \text{ yr}$, a value that is compatible with the solar rotation rate \citep{2016ApJ...819..104G}. This timescale is sufficiently short, so that any explicit effects of heat diffusion or radiative heat transfer can be omitted in global energetic balance considerations \citep{2013ApJ...777L..29C}. The boundary conditions are impermeable and stress-free for the velocity field and stop the radial flux of potential temperature perturbations. The magnetic field is assumed to be entirely radial at both boundaries.

\subsection{Ambient State}

The ambient state is computed for the polytropic stratification of an ideal gas under hydrostatic equilibrium,

\begin{align}
\frac{\partial T_{\rm e}}{\partial r} =& -\frac{g}{R (m+1)}\;,\\
\frac{\partial \rho_{\rm e}}{\partial r} =& -\frac{\rho_{\rm e}}{T_{\rm e}}
     \left(\frac{g}{R} - \frac{\partial T_{\rm e}}{\partial r} \right)\;,
\end{align}

\noindent where $R$ is the specific gas constant and $m(r)$ is the polytropic index function, set to force entropy gradients in the model. A full review of this ambient state is presented by \cite{2013ApJ...779..176G}.\\
\indent Previous works with EULAG-MHD \citep{2016ApJ...819..104G, 2019ApJ...880....6G} have shown the effect that the tachocline has on global solar models. This layer of shear at the radiative boundary begins to play a fundamental role in the storage and generation of the global magnetic field, going as far as significantly altering the time-scale of the solar cycle by decades. Another interesting development in these simulations is the formation of a near-surface shear layer (NSSL) in models with different rotational periods. In these previous works, the characteristics of convective motions were solely defined by this rotational period.\\
\indent We are interested in forcing the formation of a NSSL in an effort to study the global impacts of this boundary layer. We test several simple functions of thermodynamic ambient states generating diverse regimes of near-surface convection. Simple models of convective acceleration and deceleration offer us a chance to explore the contribution of near-surface stratification to the global distribution of angular momentum and the evolution of magnetic solar cycles. To that end we test the three following models (simulations ns1, ns2 and ns3), each having three distinct global layers: the radiation and convection zones, and the near-surface boundary layer. This is done using a polytropic index function with two step functions and three polytropic indices ($m_{\rm rz}$, $m_{\rm cz}$, and $m_{\rm ns}$), respectively representing the aforementioned layers;

\begin{equation}\label{eq:polystep}
  m(r) = m_{rz} + \dfrac{ m_{cz} - m_{rz}}{2}\left[1 + \text{erf}\left(\dfrac{r-r_{tac}}{w_{t}}\right)\right] + \dfrac{ m_{ns} - m_{cz}}{2}\left[1 + \text{erf}\left(\dfrac{r-r_{ns}}{w_{ns}}\right)\right]\;.
\end{equation}

\begin{figure}[!htb]
\centering
\includegraphics[width=\textwidth]{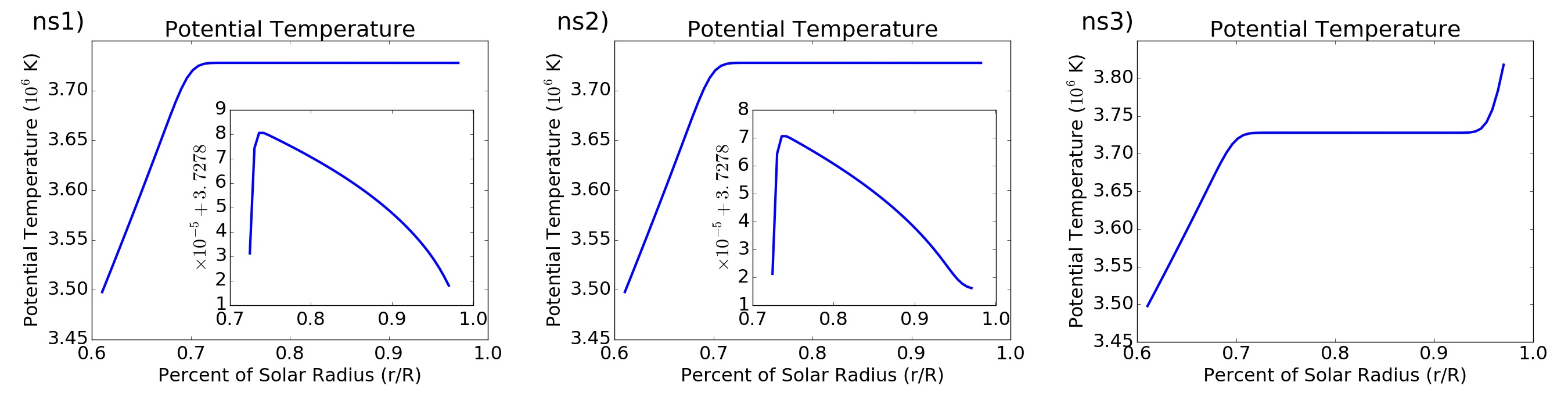}
\caption{Profiles of the ambient state of the potential temperature ($\Theta_{e}$) for simulations ns1, ns2, and ns3. In ns1, $m_{ns1} = 1.499975$, slightly increasing convection near the surface. In ns2, $m_{ns2} = 1.5$, the super-adiabatic zone is slightly altered to decrease the rate of convection. Finally, in ns3, $m_{ns3} = 1.8$. In this case convection is suppressed at the surface.}
\label{fig:ptempns}
\end{figure}

\indent The radial position of the tachocline is set at $r_{tac} = 0.72 R_{\odot}$ and the transition width is $w_{t} = 0.015 R_{\odot}$. We set our radiative interior index to $m_{rz} = 2$, creating a strongly stable subadiabatic layer; the convection zone index is set to $m_{cz} = 1.499978$ to simulate a slightly superadiabatic convective envelope. The near-surface step is centered at $r_{ns} = 0.95 R_{\odot}$ with a transition width of $w_{ns} = 0.015 R_{\odot}$. We test three different near-surface convection profiles with polytropic indeces: $m_{ns1} = 1.499975$, $m_{ns2} = 1.5$, and $m_{ns3} = 1.8$. The graph of the resulting potential temperature profiles for the three ambient states is shown in Figure \ref{fig:ptempns}.

\indent In the first model (ns1), an ambient state of slightly increased convective strength was implemented, enhancing less rotationally constrained motions near the surface and inducing the formation of a well defined near-surface shear layer. The second case, ns2, exhibits a surface entropy gradient of a hydrostatic polytrope at the edge of a stable non-convective profile ($m = 1.5$). This profile is not strong enough to fully suppress convection, but it does decrease the velocity gradient of flows near the surface. In the last simulation, ns3, a stable subadiabatic layer is created, decreasing convective motions in a low density region. This mimics a vacuum boundary for the magnetic field near the surface; a regime previously tested by \citet{2013ApJ...778..141W}.  The models are all started with an unmagnetized ambient state and are initiated with the same set of random white noise as well as perturbations of potential temperature and divergence-free velocity. We let the models evolve until the prognostic quantities achieve statistically stable equilibrium.

\section{Results}\label{sec:results}

\begin{figure}[!htb]
\flushright
\includegraphics[width=17cm]{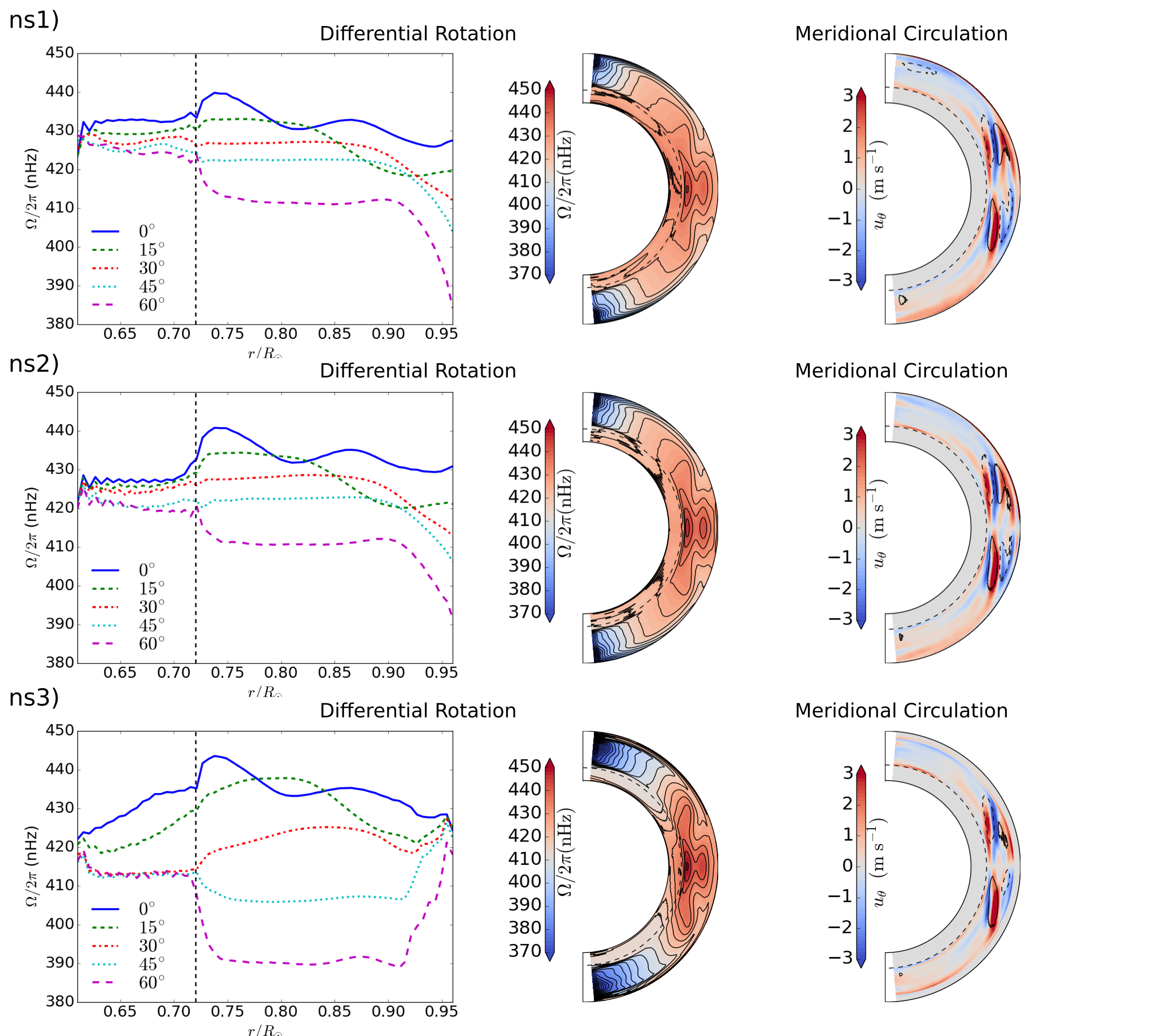}
\caption{The time averaged (over three dynamo cycles) and azimuthally-averaged differential rotation and meridional circulation profiles for ns1, ns2, and ns3, from top to bottom. On the left-hand side we see the angular velocity at various latitudes as a function of radius. In the center we show a contour plot of this differential rotation in the meridional plane. On the right we show the meridional circulation in the same plane, where colored contours show the speed of latitudinal flows; the dashed and solid contours outline clockwise and counter-clockwise poloidal circulation.}
\label{fig:hydro}
\end{figure}

\begin{figure}[!htb]
\flushleft
\includegraphics[width=17cm]{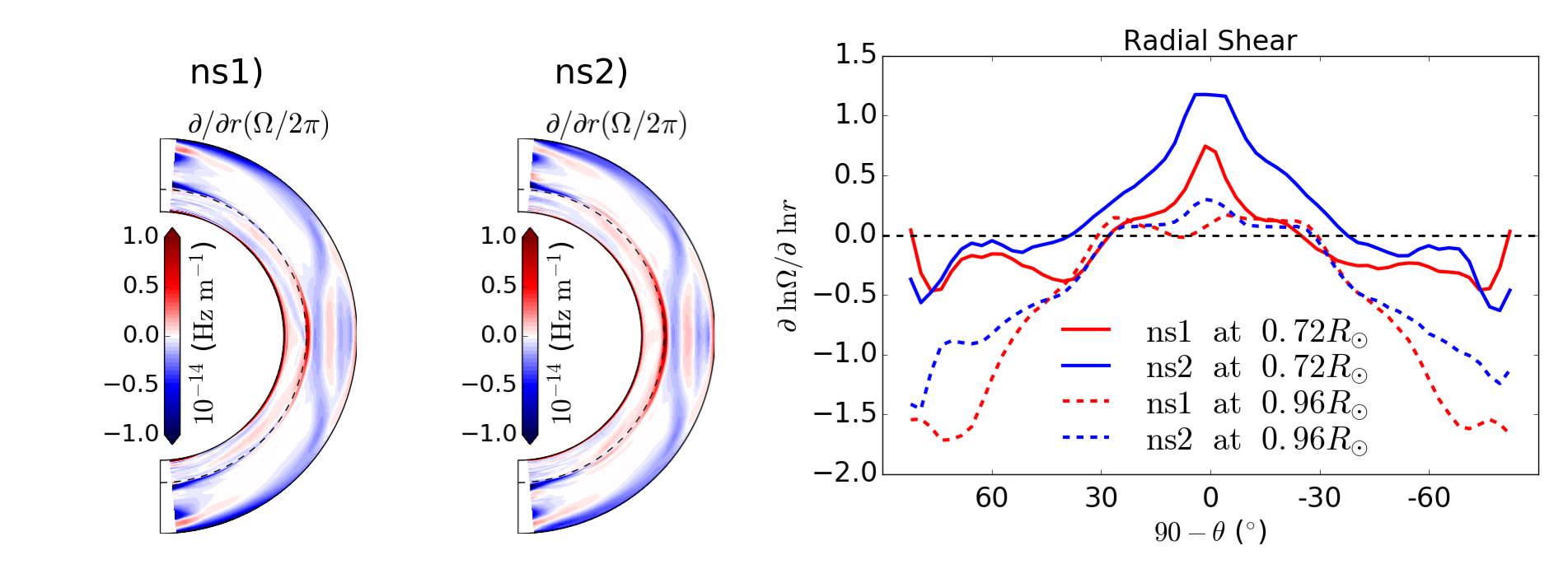}
\caption{Radial shear profiles of simulations ns1 (left panel) and ns2 (middle) characterized by the radial derivative of the angular velocity. The largest noticeable differences are of the shear experienced at the tachocline and at the poles near the surface (right), an effect consistent with the observed differences in the models' rate of angular momentum distribution into the radiation zone (Fig. \ref{fig:hydro}).}
\label{fig:om_rdv}
\end{figure}

\indent We let each simulation evolve for several hundred years and proceed to analyze the last three polarity reversals with an oscillatory period of the large-scale magnetic field of $\sim 30\text{ yrs}$ for ns1, $\sim 20 \text{ yrs}$ for ns2, and $\sim 40\text{ yrs}$ for ns3. Each model develops unique rotational hydrodynamic structures and global magnetic profiles; in Figure \ref{fig:hydro}, we present the time-averaged (over three magnetic cycles) and azimuthally-averaged differential rotation profile along with the meridional circulation.\\
\indent All three simulations exhibit profiles which appear to be similar to the solar model (RC02), discussed by \citet{2016ApJ...819..104G}; there are, however, a few notable differences. The effect of suppressed surface convection is immediately obvious in the velocity profile of ns3, where large differences with the other two models begin to develop. A buildup of rotational velocity near the equator can be observed above the tachocline, with less angular momentum being carried into the radiative interior as well as being excluded from the surface. This model also exhibits strong latitudinal variance of rotational frequency, displaying significant gradation in the differential rotation profile -- extending well into the radiative interior. The sub-surface region of simulation ns3 manifests a strong positive shear, in contrast with the other two models, ns1 and ns2. The most significant differences are found below the tachocline, where angular momentum transfer fails to penetrate, especially at higher latitudes, and maintains the models' initial rotation rate ($\Omega/2\pi \approx 413$ nHz).\\
\indent The models of slightly increased (ns1) and suppressed (ns2) convection rates, predictably, look similar, with certain exceptions. The rotation rate of the radiative interior exhibits the greatest differences between the two models, with simulation ns1 reaching an average angular frequency of $\sim 430\text{ nHz}$ as compared to $\sim 425\text{ nHz}$ of simulation ns2. Significant differences can also be observed in the rate of rotational frequency observed at higher latitudes. Even though these results do not closely match helioseismological inferences of internal solar rotation rates \citep{1998ApJ...505..390S}, the models do show clear conical iso-contours as opposed cylindrical ones, breaking the Taylor-Proudman balance at the upper latitudes. Rotational frequency in the convection zone (Figure \ref{fig:hydro}), however, still shows a shallow break in differential rotation near the equator, aligned along the rotational axis.\\
\indent One of the largest effects of the sub-surface acceleration can be seen in its impact on the levels of shear experienced at the tachocline (Fig. \ref{fig:om_rdv}), decreasing positive levels of shear near the equator and increasing negative shear near the poles. The result is a changing focus of the $\Omega$-effect, concentrating greater action into higher latitudes of model ns1.

\begin{figure}[!htb]
\centering
\includegraphics[width=9.5cm]{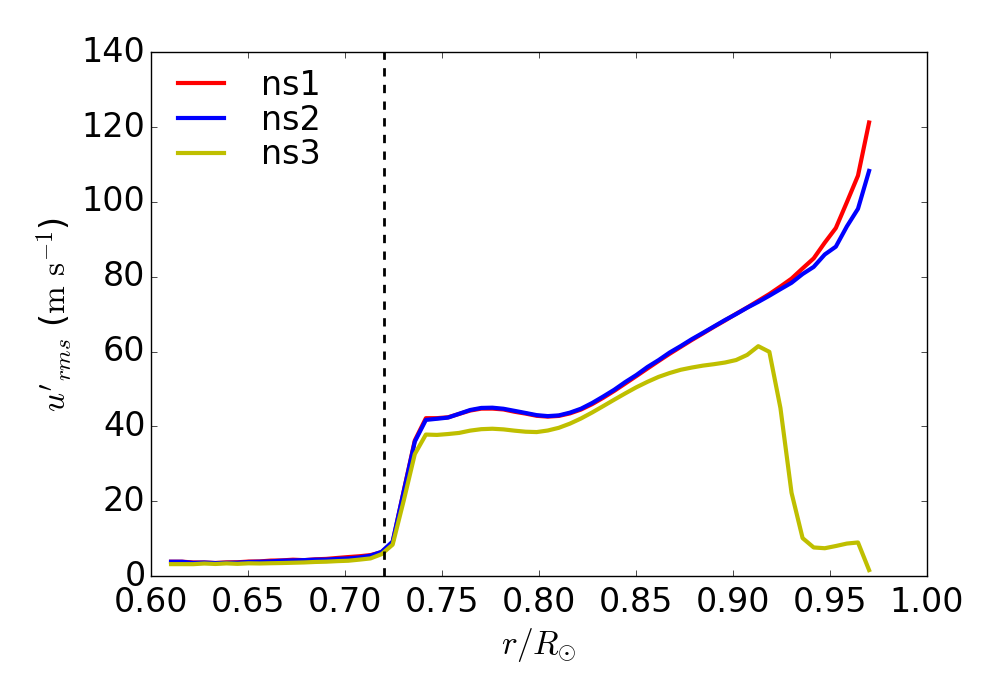}
\caption{Radial profile of the rms velocity (${u'_{rms}}$). Models ns1 and ns2 are very similar with increasing turbulence near the surface. In model ns3 the subadaiabtic gradient effectively quenches strong turbulence deep into the convection zone.}
\label{fig:urms}
\end{figure}

\indent The shear profiles of these models (ns1, ns2) are structurally similar to those expected from helioseismological inferences \citep{1998ApJ...505..390S}, with a few significant differences. The bulk of the convection zone contains contours of shear at low latitudes - matching the axis-aligned contour in differential rotation, breaking the conical differential rotational cell seen in Figure \ref{fig:hydro}. This is most likely a consequence of the models limited ability to fully break the Taylor-Proudman balance. Near the model surface we also see some deviation from observed solar rotation where a consistent shear ($\partial\ln\Omega/\partial\ln r = -1$) has been observed up to $\sim60^{\circ}$ latitude \citep{2014A&A...570L..12B}. Even though near-surface shear in the ns1 and ns2 models hovers around this value it begins to diverge at low latitudes, becoming positive (see right panel of  Fig.~\ref{fig:om_rdv}). Such an acceleration has not been observed in recent hydrodynamic simulations of near-surface gradation \citep{2019ApJ...871..217M}, where different density contrasts are modeled in an attempt to generate solar-like near-surface shear. Their model with large density stratification results in negative surface shear at equatorial latitudes. Our models ns1 and ns2, in comparison, implement various potential temperature contrasts in the attempt to create rotationally unconstrained flows. Increasing the density contrast might allow this region to break up the axis-aligned structures of shear and induce the development of negative radial shear at equatorial latitudes. Our models also do not consider the top $3\%$ of the solar surface -- a turbulent region of a large density stratification where the time-scale of convective motions is considerably shorter than that of solar rotation.\\
\indent An interesting feature is exhibited in model ns1, where the rotation of the radiation zone is not completely hemispherically symmetric, unlike the two others - a possible structural consequence of the differences in the development of their respective dynamo profiles. This effect is also apparent in the distribution of shear (Fig. \ref{fig:om_rdv}), which shows a small asymmetrical tilt towards the northern hemisphere.\\
\indent All three models exhibit a meridional circulation pattern observed in previous EULAG-MHD models \citep{2016ApJ...819..104G}. The only noticeable difference between the models is a slight poleward increase in meridional velocity at higher latitudes, primarily on the surface of model ns1. These meridional profiles show the formation of a two cell structure, recently observed by helioseismology \citep{2013ApJ...774L..29Z}; these cells are, however, confined to low latitudes and aligned with the rotational axis - similar profiles have been previously observed in other global anelastic models \citep[e.g.,][]{2002ApJ...570..865B, 2015ApJ...804...67F, 2019ApJ...871..217M}.\\
\indent To better understand the resulting turbulent structure of our models we plot the perturbative rms velocity (${u'_{rms}} = \sqrt{u'_{r} + u'_{\theta} + u'_{\phi}}$) in Figure~\ref{fig:urms}. In model ns3, we observe a significant drop in turbulent velocity near the surface, consistent with the strong subadiabatic gradient that we induced. In models ns1 and ns2, the differences are rather small with only the obvious slight changes near the surface. The turbulent ${u'_{rms}}$ is largely isotropic in these models, with the exception of the region near the surface where the boundary condition transforms the radial velocity (and its perturbative component) into horizontal flows.

\begin{figure}[!htb]
\centering
\includegraphics[width=\textwidth]{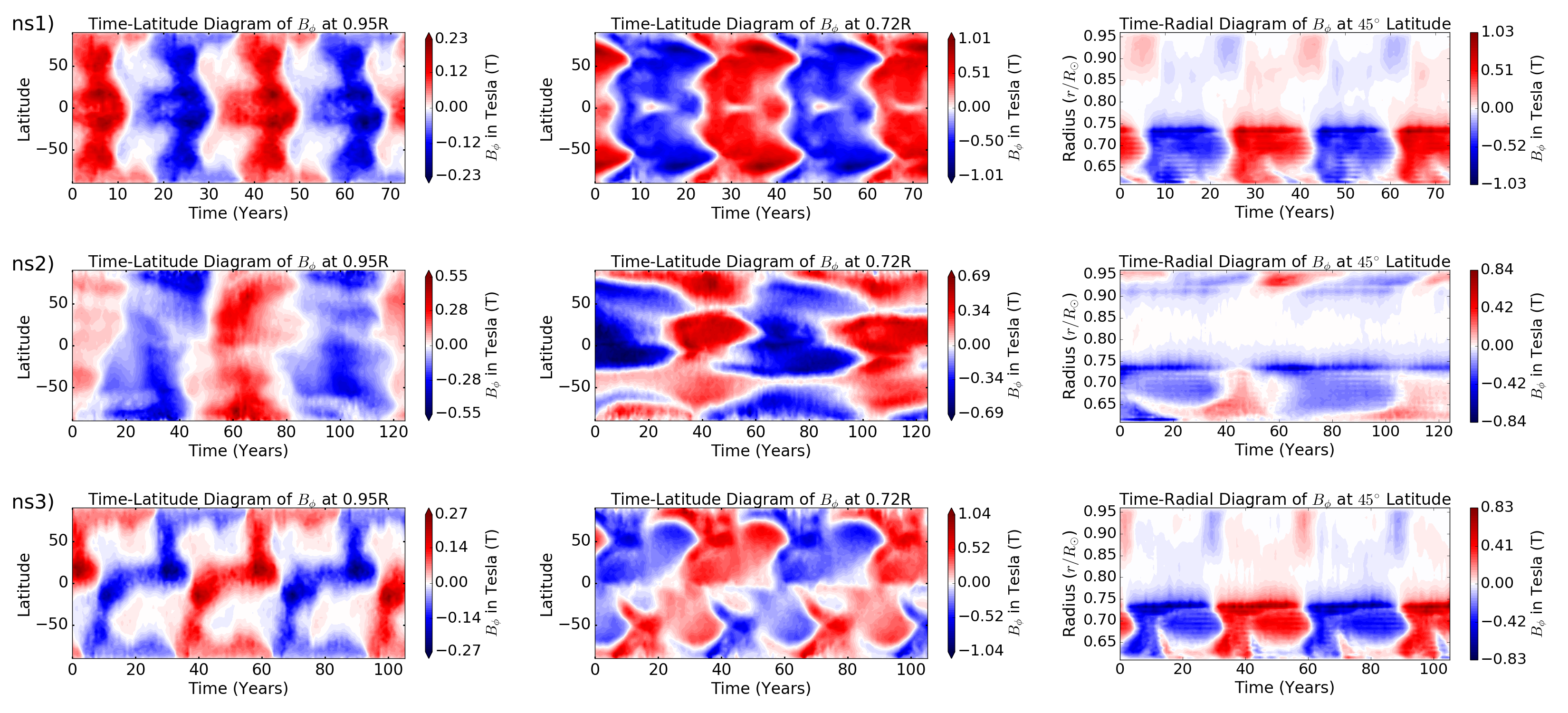}
\caption{The time-latitude diagrams of the magnetic field ($B_{\phi}$ measured in Tesla) for simulations ns1, ns2, and ns3 (from top to bottom) at 0.95 $R_{sun}$ (on the left) and 0.72 $R_{\odot}$ (at the center), as well as the time-radius  diagram at $45^{\circ}$ latitude (on the right). ns1) With slightly increased convection, the beginnings of a staggered anti-symmetric magnetic cycle start to form. ns2) The field follows a distinct pattern, but the anti-symmetric nature we observe on the Sun is lost. ns3) The convection profile is altered such that the field is unstable and non-uniform, but with a well defined cyclic pattern.}
\label{fig:mag}
\end{figure}

\indent These relatively small differences in convective structures result in large impacts on emerging mean flows and global magnetic fields. The time evolution of the magnetic dynamo over the course of three cycles is shown in Figure~\ref{fig:mag}, plotting a cross section of the azimuthally averaged toroidal magnetic field, $\overline{B}_{\phi}$, near the model surface ($\sim 0.95 R_{\odot}$), at the tachocline ($\sim 0.72 R_{\odot}$), as well as a radial cross section at a latitude of $45^{\circ}$.

\indent Model ns3 exhibits the largest difference in its profile of polarity reversals, where the pattern of the global magnetic field cannot seem to find a definite symmetric hemispheric structure as well as exhibiting the longest periodic evolution of its dynamo cycle ($\sim 40\text{ yrs}$). Even in the absence of turbulent activity in the upper convection zone, a strong magnetic field is still being generated and stored, loosely connected with strong fields below the tachocline. These fields still undergo regular polarity reversals - but without any defined pattern or symmetric structure. The models with slightly suppressed (ns2) and slightly increased convection (ns1) however, both exhibit regular cyclic patterns, with large differences in the time-scale of their evolution as well as the nature of their symmetries. The toroidal field profile of model ns2 looks similar to the solar model RC02 (with no alterations of the sub-surface boundary) described by \cite{2016ApJ...819..104G}, showing minimal structural effects of its slightly suppressed convective motions. The most significant difference appears to be with the increased convection model (ns1), where the emerging global magnetic field begins to shift from the equatorial symmetry seen in profile ns2 to an offset, staggered near anti-symmetry, exhibiting a more solar-like hemispheric polarity.\\

\subsection{Mean-Field Analysis}

\indent For an in depth analysis of the dynamo characteristics that result in the large structural differences in models ns1 and ns2, we rewrite our induction equation (Eq. \ref{eq:gov4}) into its mean-field form \citep{1978mfge.book.....M},

\vskip-0.1cm
\begin{align}\label{eq:mfb}
    \dfrac{\partial\mathbf{\overline{B}}}{\partial t} = \nabla\times(\mathbf{\overline{u}}\times\mathbf{\overline{B}} + \boldsymbol{\overline{\mathcal{E}}})\;, && \text{where,\indent} \boldsymbol{\overline{\mathcal{E}}} = \overline{\mathbf{u'}\times\mathbf{B'}}\;.
\end{align}

\indent The mean-field terms (denoted by the overline) are averaged over their longitude, with primed terms being the perturbation from this average. Using the first order smoothing approximation (FOSA), we rewrite our turbulent induction, $\boldsymbol{\overline{\mathcal{E}}}$ (excluding any triple correlation terms and third order derivatives), in the manner of \cite{2005PhR...417....1B}, with the turbulent transport coefficients defined as

\begin{figure}[!htb]
\centering
\includegraphics[width=\textwidth]{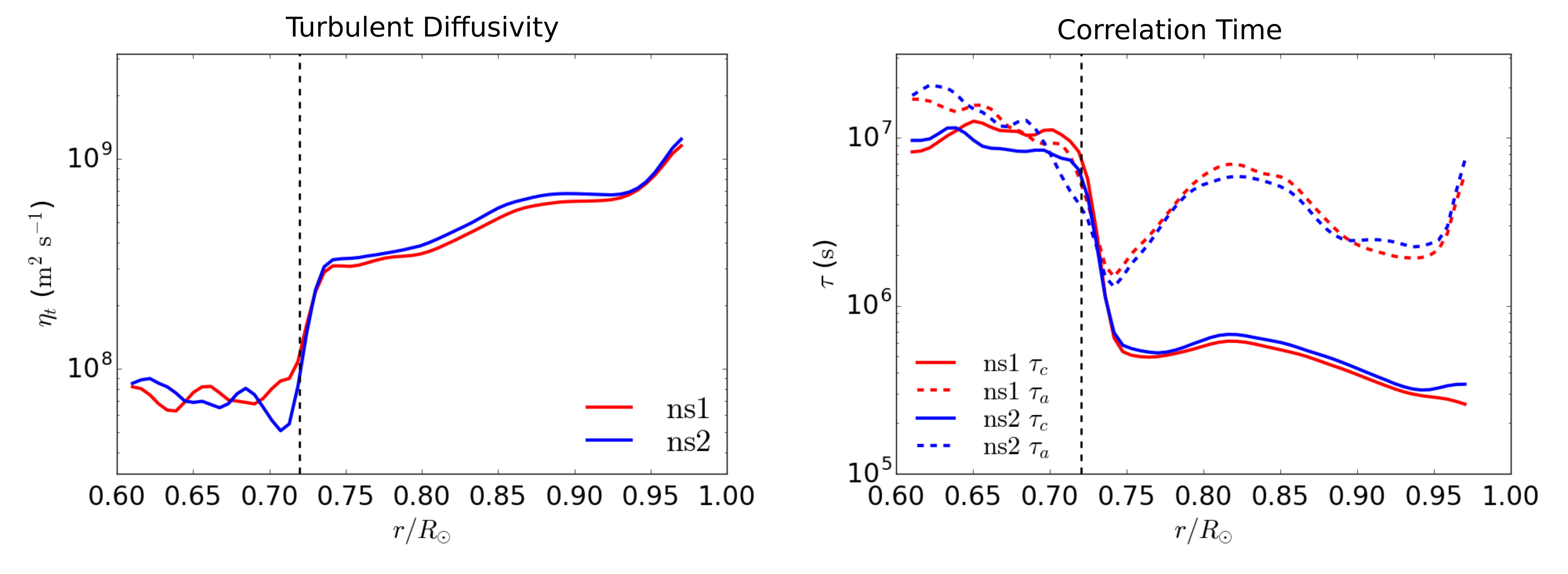}
\caption{Left: radial dependence of the turbulent diffusivity, $\eta_t$, and right: the Alfv\'en (dashed lines) and kinetic (solid) correlation times, $\tau_{a}$ and $\tau_{k}$, as defined by the first order smoothing approximation (FOSA, Eq. \ref{eq:fosa}).}
\label{fig:eta_tau}
\end{figure}

\vskip-0.2cm
\begin{align}\label{eq:fosa}
    \overline{\mathcal{E}}_{i} = (\alpha_{k} + \alpha_{m}) \overline{B}_{j} - \eta_{t} \frac{\partial \overline{B}_{j}}{\partial x_{k}} \; ; &&
    \begin{aligned}
        \alpha_{k} &= -\frac{1}{3}\tau_{corr}(\overline{\boldsymbol{\omega'}\cdot\mathbf{u'}})\; ,\\
        \alpha_{m} &= \frac{1}{3}\tau_{corr}(\overline{\mathbf{j'}\cdot\mathbf{B'}}) \; ,\\
        \eta_{t} &= \frac{1}{3}\tau_{corr}(\mathbf{\overline{u'^{2}}})\; ,
    \end{aligned}
\end{align}

\noindent where $\boldsymbol{\omega'} = \nabla\times\mathbf{u'}$ is the turbulent vorticity and $\mathbf{j'} = \nabla\times\mathbf{B'}$ is a perturbation of the induced current. $\tau_{corr}$ is defined as the correlation time of turbulent motions. To efficiently compute this correlation time, we adopt the spectral approach outlined by \citet{2019ApJ...880....6G}, using turbulent energy spectra to compute the integral length scale $\left(l(r) = r \int E(r,k)k^{-1} dk/\int E(r,k) dk\right)$ as a function of radius. This approach breaks down, however, in the region below the tachocline ($r<0.72R_{\odot}$) where convective turbulence is effectively quenched (Figures \ref{fig:urms}, \ref{fig:eta_tau}). We can employ the Alfv\'en velocity of strong magnetic fields stored in the region to compute the correlation time of the current helicity. Two length scales ($l_{k}, l_{m}$) are computed from their respective kinetic and magnetic energy spectra ($E(u')$, $E(B')$) and the resulting correlation times are calculated as follows: $\tau_{k} = l_{k}/u'_{rms}$, $\tau_{a} = l_{m}/v_{a}$, where the Alfv\'en velocity is defined as $v_{a} = B'_{rms}/\sqrt{\mu_{0}\rho_{e}}$. The resulting correlation times are presented on the right hand side of Figure \ref{fig:eta_tau}.

\begin{figure}[h]
\centering
\includegraphics[width=16cm]{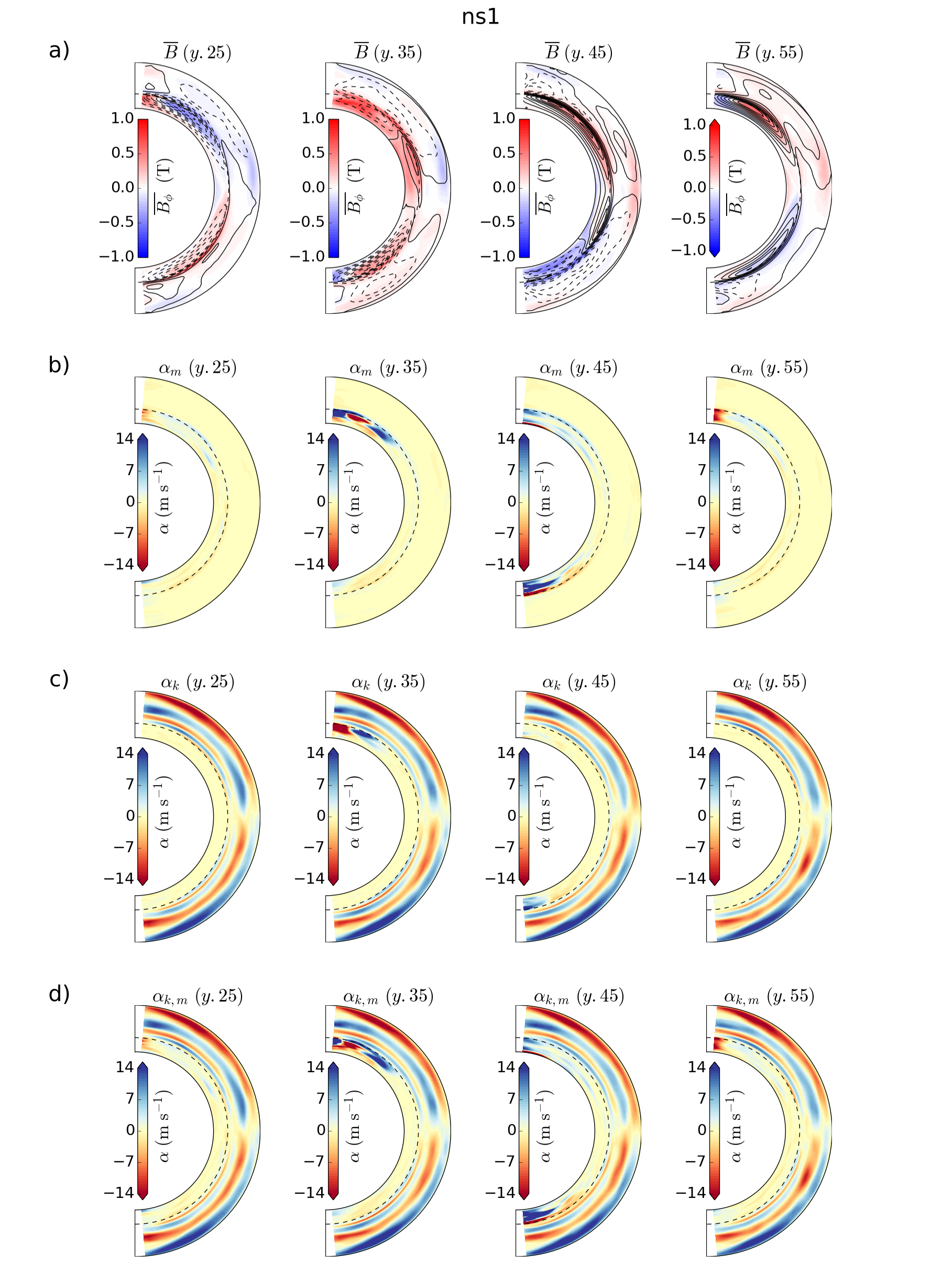}
\caption{Snapshots of polarity reversal for model ns1. a) The toroidal magnetic field ($\overline{B_{\phi}}$) exhibited by the colored contour with the colorbar; the poloidal magnetic field ($\overline{B_{p}}$) drawn by the solid/dashed contour lines. b) The magnetic turbulent transport coefficient $\alpha_{m}$. c) The kinetic turbulent transport coefficient $\alpha_{k}$. d) A summation of the two coefficients.}
\label{fig:alpha3}
\end{figure}

\begin{figure}[h]
\centering
\includegraphics[width=16cm]{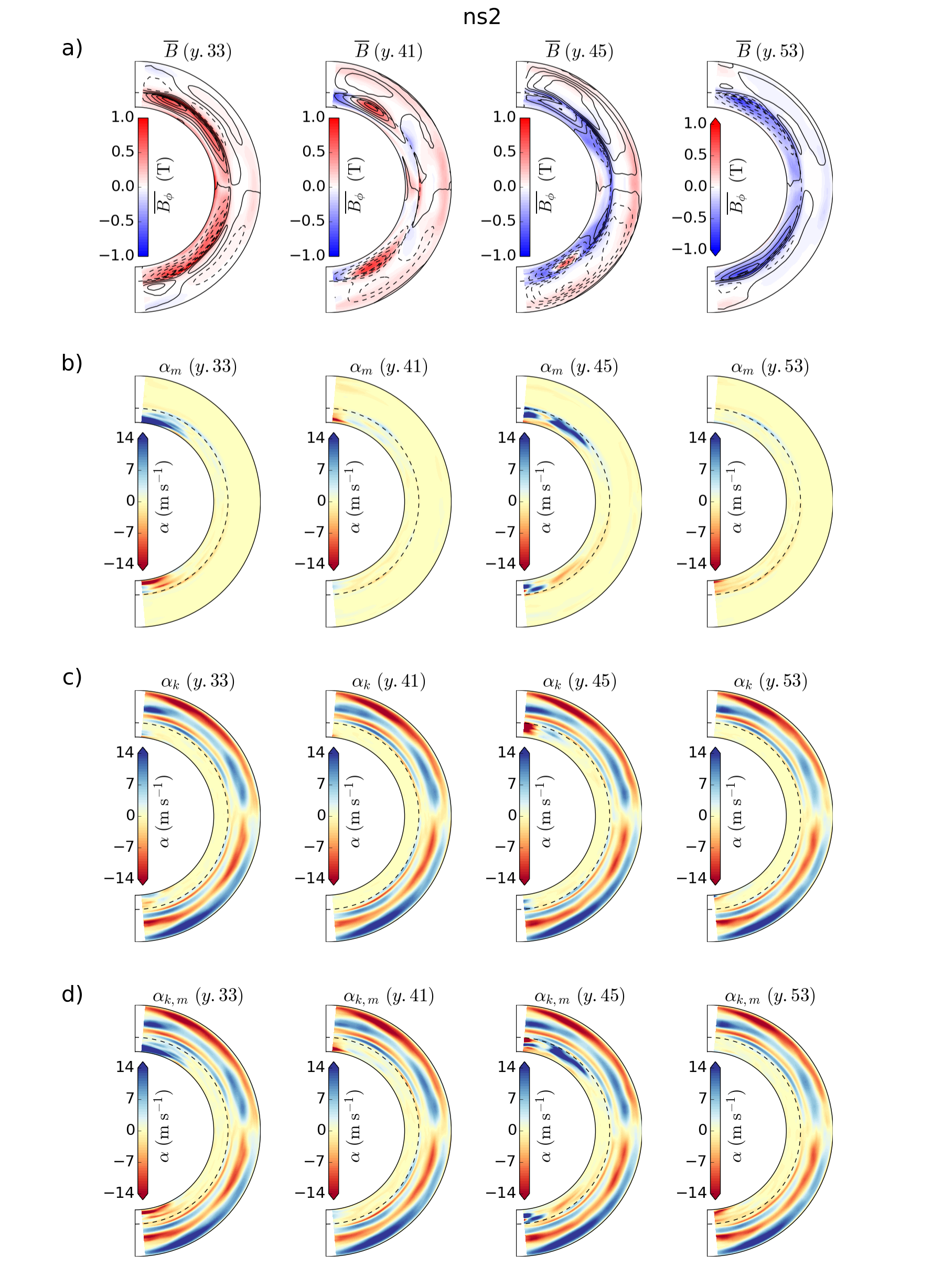}
\caption{Same as Figure~\ref{fig:alpha3} for simulation ns2.}
\label{fig:alpha1}
\end{figure}

\indent These profiles display little indication to the cause of the evident shift in the emergence of the global magnetic field between models ns1 and ns2, with only slight differences in turbulent diffusivity ($\eta_{t}$) and correlation time ($\tau_{corr}$) in areas directly below the tachocline. To further explore the differences in the interplay of turbulent coefficients in these models, we present contour plots (Figure~\ref{fig:alpha3} and \ref{fig:alpha1}) of the global magnetic fields during their transition from one polarity to another, sampled at $t = 25$, $35$, $45$ and $55$ years for simulation ns1 and $t=33$, $41$, $45$ and $53$ years for simulation ns2, along with the corresponding levels of turbulent kinetic and magnetic $\alpha$-effects, $\alpha_{k}$ and $\alpha_{m}$; see Eq. \ref{eq:fosa}.

\begin{figure}[!htb]
\flushleft
\includegraphics[width=17cm]{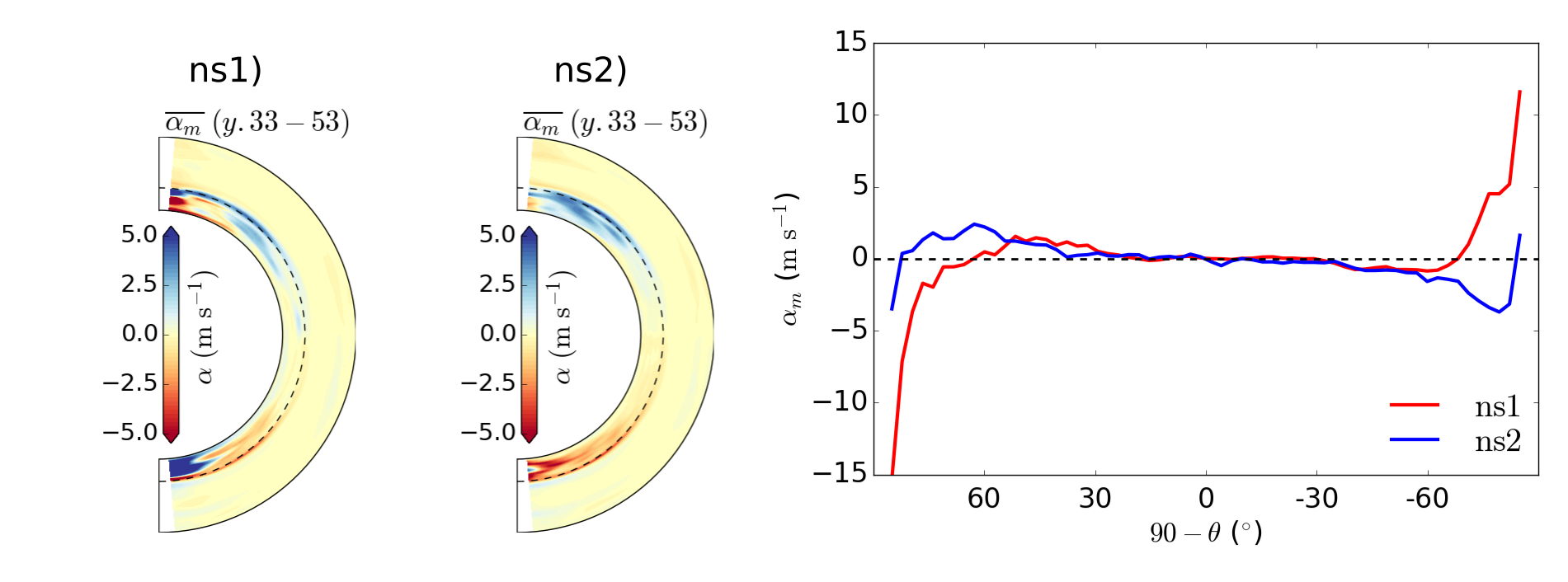}
\caption{The magnetic turbulent transport coefficient, $\alpha_{m}$, computed from the current helicity in the FOSA/MTA approximations (Eq.~\ref{eq:fosa}). There are significant differences between the models ns1 and ns2, denoted by the emergence of strong helicities near the poles of model ns1.}
\label{fig:alpha_cmp}
\end{figure}

\indent The transition snapshots (Figures~\ref{fig:alpha3}, \ref{fig:alpha1}) exhibit clear differences between the models, specifically in the asymmetry of the evolving field in model ns1 (Fig.~\ref{fig:alpha3}); the mean poloidal field, $\overline{\bf B}_{p} = (\overline{B}_{r},\overline{B}_{\theta},0)$, begins its reversal at the tachocline, extending into the radiation zone (y.35) in the region of minimal hydrodynamic shear between the radiation and convection zones (see also Fig. \ref{fig:hydro}). A strong poloidal field of a single polarity proceeds to encompass the entire radiation zone. This development is accompanied by the development of strong toroidal fields ($B_{\phi}$) under the tachocline in hemispherically anti-symmetric bands. This evolution is followed by much weaker fields in the convection zone. The convectively suppressed model (ns2, Fig. \ref{fig:alpha1}) experiences the opposite pattern of development, where a single symmetric toroidal field ($\overline{B}_{\phi}$) band encompasses both hemispheres, with two hemispherically reflected anti-symmetric poloidal field cells developing (y.45) at low latitudes at the base of the tachocline.\\
\indent The kinetic $\alpha$-effect is maintained almost entirely consistently over polarity reversals in ns1 and ns2; $\alpha_{k}$ emerges in hemispherically anti-symmetric bands near the surface, followed by a reversal at further depths. In the convection zone, $\alpha_k$ appears to dominate $\alpha_m$ by an order of magnitude. Similar turbulent transport coefficient profiles in EULAG-MHD models are explored in further detail by \citet{2019ApJ...880....6G} while the effectiveness of the first-order smoothing approximation (FOSA) is thoroughly investigated by the test-field method of \citet{2018A&A...609A..51W}. Over the course of the polarity reversals, the region underneath the tachocline experiences the greatest deviation in its pattern of activity as well as the strongest manifestation of the turbulent $\alpha$-effect. A relevant parameter for the evolution of the global magnetic field seems to be the high level of current helicity ($\alpha_{m}$) generated underneath the tachocline. Due to the large values of $\tau_a$ in the radiation zone, the magnetic $\alpha$-effect is dominant in the region, intermittently reaching levels much higher than kinetic turbulent helicity in the convection zone. These cyclonic motions are prominent during periods of transition when strong toroidal fields encompass the tachocline and radiation zone at higher latitudes. Figures~\ref{fig:alpha3} and \ref{fig:alpha1} suggest that $\alpha_m$ is the potential source of the large variations observed in the models. The aggregate effects of turbulence in this region can be seen in the average of the magnetic turbulent coefficient ($\alpha_{m}$) over these polarity reversals (Fig. \ref{fig:alpha_cmp}).

\indent These simulations see an $\alpha_{m}$ term with an opposite sign begin to emerge at the poles with a very weak, but noticeable, onset in the suppressed model ns2. This term grows to a large field source in ns1, more than three times as strong as the bands of opposing helicities extended into the lower latitudes of the radiation zone. These polar helicities are the strongest contributors to the $\alpha$-effect at and underneath the tachocline. During polarity reversals, these regions intermittently become the single strongest turbulent sources of the poloidal magnetic field in these models.

\section{Discussion}\label{sec:conclusions}

\indent The simulations presented in this paper are a continuation of previous work done with EULAG-MHD \citep{2010ApJ...715L.133G, 2013ApJ...779..176G, 2016ApJ...819..104G, 2019ApJ...880....6G}, using the resolution, 128 ($\phi$), 64 ($\theta$) and  64 ($r$) grid points to simulate solar convective rotating turbulence and dynamo action in the numerical viscosity regime.\\
\indent In the study of solar convection, the results of our simulations ns1, ns2 and ns3 illustrate the relevance of the stratification in the near-surface layer. Recent simulations done by \cite{2019SciA....5.2307H} (sans rotation and magnetic field) show that the near-surface region does not affect the amplitude of the convective motions in the deep interior. Our results, however, suggest that, for rotating hydro-magnetic convection, small convective acceleration close to the surface was sufficient to impact the redistribution of angular momentum, especially into the radiative zone. For simulations ns1 and ns2 there is a marked increase in the levels of rotational frequency, while for simulation ns3, where the surface flows were quenched, the rotational frequency of the radiation zone remained the same as that of the reference frame (especially at higher latitudes).\\
\indent The largest hydrodynamic impact at the tachocline is a movement away from the strongest velocity gradients found at the equator in simulation ns2, along with the solar rotation model RC02 in \cite{2016ApJ...819..104G}, to stronger velocity gradients as we approach the poles in simulation ns1 (Fig. \ref{fig:om_rdv}). In ns1, the radiative layer reaches closer to the angular frequency of the convection zone near the equator, similar to the helioseismological inferences of solar observations \citep{1998ApJ...505..390S}.\\
\indent These models develop dynamo cycles rooted in the area underneath the tachocline ($r < 0.72R_{\odot}$), with strong magnetic fields generated and stored in the radiation zone (Fig. \ref{fig:mag}). The magnetic fields in these regions undergo a much lower rate of turbulent diffusivity (Fig.~\ref{fig:eta_tau}), extending the lifetime of their cycle. High levels of $\alpha_{m}$ and shear at the tachcoline (Fig. \ref{fig:om_rdv}) allow these regions to become strong sources of poloidal/toroidal transitions (Figures~\ref{fig:alpha1} and \ref{fig:alpha3}).\\
\indent The source of variance in hemispheric parity between models ns1 and ns2 (Fig. \ref{fig:mag}) is not entirely clear; all of the model's parameters, except for the near-surface boundary, are identical. There are, however, notable differences in the development of tachoclinic shear (Fig. \ref{fig:om_rdv}) as well as substantial differences in the development of current helicity, and therefore $\alpha_{m}$ (Fig.~\ref{fig:alpha_cmp}) -- appearing as emerging turbulent sources near the poles of an opposing sign to the helicity that covers the rest of the hemisphere. These differences occur in regions of large hydrodynamic variations, exhibited by increasing rates of rotational frequency (Fig. \ref{fig:hydro}) from ns2 to ns1 ($\sim 425\text{ nHz}$ to $\sim 430\text{ nHz}$) and evident when comparing rotation rates near the poles; acceleration appears to filter up the latitudes in concert with increased convection rates near the surface.\\
\indent The source of the strong current helicities and their orientation is not yet entirely clear, but, it is notable that they are generated directly over the poles, where the Coriolis force has little influence on the orientation of the turbulent vorticity. It is also telling that these helicities are more strongly generated in a regime of increased polar downflow, following the increased sub-surface convection in simulation ns1. As the majority of the impact of turbulent coefficients is shifted towards higher latitudes, they may result in a more stable manifestation of the dipolar global magnetic dynamo.

\acknowledgments

\indent AMS would like to thank the team at NASA Ames Research Center for their help and expertise, as well as the referee for his/her constructive comments. This work is supported by the NASA grants: 80NSSC19K0630, 80NSSC19K1436, NNX14AB7CG and NNX17AE76A.


\begin{thebibliography}{}


\bibitem[Augustson et al.(2015)]{2015ApJ...809..149A} Augustson, K., Brun, A.~S., Miesch, M., Toomre, J.\ 2015, \apj, 809, 149 

\bibitem[Babcock(1961)]{1961ApJ...133..572B} Babcock, H.~W.\ 1961, \apj, 133, 572

\bibitem[Barekat et al.(2014)]{2014A&A...570L..12B} Barekat, A., Schou, J., Gizon, L.\ 2014, \aap, 570, 12

\bibitem[Bonanno et al.(2002)]{2002A&A...390..673B} Bonanno, A., Elstner, D., Rüdiger, G., Belvedere, G.\ 2002, \aap, 390, 673

\bibitem[Brandenburg(2005)]{2005ApJ...625..539B} Brandenburg, A.\ 2005, \apj, 625, 539

\bibitem[Brandenburg \& Subramanian(2005)]{2005PhR...417....1B} Brandenburg, A., Subramanian, K.\ 2005, Physics Reports, 417, 1

\bibitem[Brun \& Toomre(2002)]{2002ApJ...570..865B} Brun A.~S., Toomre J.\ 2002, \apj, 570, 865

\bibitem[Brun et al.(2004)]{2004ApJ...614.1073B} Brun A.~S., Miesch M.~S., Toomre J.\ 2004, \apj, 614, 1073

\bibitem[Dikpati \& Charbonneau(1999)]{1999ApJ...518..508D} Dikpati, M., Charbonneau, P.\ 1999, \apj, 518, 508

\bibitem[Charbonneau(2010)]{2010LRSP....7....3C} Charbonneau, P.\ 2010, LRSP, 7, 3 

\bibitem[Chatterjee et al.(2004)]{2004A&A...427.1019C} Chatterjee, P., Nandy, D., Choudhuri, A.~R.\ 2004, \aap, 427, 1019

\bibitem[Cossette et al.(2013)]{2013ApJ...777L..29C} Cossette, J., Charbonneau, P., Smolarkiewicz, P.~K.\ 2013, \apj, 777, 29 

\bibitem[Cossette et al.(2017)]{2017ApJ...841...65C} Cossette, J., Charbonneau, P., Smolarkiewicz, P.~K., Rast M.P., \ 2017, \apj, 841, 65

\bibitem[Dikpati \& Gilman(2001)]{2001ApJ...559..428D}  Dikpati, M., Gilman. P.~A.\ 2001, \apj, 559, 428 

\bibitem[Domaradzki et al.(2003)]{Domar03} Domaradzki, J.A., Xiao, Z. Smolarkiewicz, P.K.\ 2003, Phys. Fluids, 15, 3890 

\bibitem[Elliott \& Smolarkiewicz(2002)]{elliott02} Elliott, J.R., Smolarkiewicz, P.K.\ 2002, Int. J. Numer. Methods Fluids, 39, 855 

\bibitem[Featherstone \& Miesch(2015)]{2015ApJ...804...67F} Featherstone, N.~A., Miesch M.~S.\ 2015, \apj, 804, 67
	
\bibitem[Germano et al.(1991)]{germano91} Germano, G., Piomelli, U., Moin, P., Cabot, W.~H.\ 1991, Phys. Fluids A: Fluid Dynamics, 3, 1760 

\bibitem[Ghizaru et al.(2010)]{2010ApJ...715L.133G} Ghizaru, M., Charbonneau, P., Smolarkiewicz, P.~K.\ 2010, \apjl, 715, 133 

\bibitem[Gilman \& Miller(1981)]{1981ApJS...46..211G} Gilman, P.~A., Miller, J.\ 1981, ApJS, 46, 211 

\bibitem[Grinstein et al.(2007)]{iles07} Grinstein, F. F., Margolin, L. G., \& Rider, W. J.\ 2007, Cambridge University Press 

\bibitem[Guerrero et al.(2008)]{2008A&A...485..267G} Guerrero, G., Gouveia Dal Pino, E.~M.\ 2008, \aap, 485, 267 

\bibitem[Guerrero et al.(2013)]{2013ApJ...779..176G} Guerrero, G., Smolarkiewicz, P.~K., Kosovichev, A.~G., Mansour N.~N.\ 2013, \apj, 779, 176

\bibitem[Guerrero et al.(2016)]{2016ApJ...819..104G} Guerrero, G., Smolarkiewicz, P.~K., Gouveia Dal Pino, E.~M., Kosovichev, A.~G., Mansour N.~N.\ 2016, \apj, 819, 104 

\bibitem[Guerrero et al.(2019)]{2019ApJ...880....6G} Guerrero, G., Zaire, B., Smolarkiewicz, P.~K., et al.\ 2019, \apj, 880, 6

\bibitem[Hotta (2019)]{2019SciA....5.2307H}Hotta, H.; Iijima, H.; Kusano, K.\ 2019, SciA, 5, 2307

\bibitem[Karak et al.(2014)]{2014ApJ...795...16K} Karak, B.~B., Rheinhardt, M., Brandenburg, A., K\"apyl\"a, P.~J., K\"apyl\"a, M.~J. \ 2014, \apj, 795, 16

\bibitem[K\"uhnlein et al.(2019)]{kuhnlein19} K\"uhnlein, C., Deconinck, W., Rupert, K., et al.\ 2019, Geosci. Model Dev., 12, 651 

\bibitem[Kumar et al.(2015)]{Kumar15} Kumar S., Bhattacharyya, R. \& Smolarkiewicz, P.~K.\ 2015, Phys. Plasmas, 22, 082903 

\bibitem[Kosovichev et al.(1997)]{kosovichev97} Kosovichev, A.~G., Schou, J., Scherrer, P.~H., et al.\ 1997, \solphys, 170, 43 

\bibitem[Leighton(1969)]{1969ApJ...156....1L} Leighton, R.~B.\ 1969, \apj, 156, 1
 
\bibitem[Lilly(1966)]{lilly66}Lilly, D.~K.\ 1966, NCAR Manuscript No. 123 

\bibitem[Lipps \& Hemler(1982)]{LH82} Lipps, F.~B., Hemler, R.~S.\ 1982, JAS, 39, 2192 

\bibitem[Matilsky et al.(2019)]{2019ApJ...871..217M} Matilsky, L.~I.; Hindman, B.~W., Toomre, J.\ 2019, \apj, 871, 217

\bibitem[Margolin et al.(2002)]{margolin02} Margolin, L.~G., Smolarkiewicz, P.~K., Wyszogrodzki, A.~A.\ 2002, J. Fluids Eng., 124, 862 

\bibitem[Margolin et al.(2006)]{margolin06} Margolin, L.~G., Smolarkiewicz, P.~K., Wyszogrodzki, A.~A.\ 2006, J. Appl. Mech., 73, 469  

\bibitem[Moffatt(1978)]{1978mfge.book.....M} Moffatt, H.~K.\ 1978, Cambridge University Press, 353 

\bibitem[Nandy \& Choudhuri(2002)]{2002Sci...296.1671N} Nandy, D., Choudhuri, A.~R.\ 2002, \aap, 296, 1671

\bibitem[Nelson et al.(2013)]{2013ApJ...762...73N} Nelson, N.~J., Brown, B~.P., Brun, A.~S., Miesch, M.~S., Toomre, J.\ 2013, \apj, 762, 73
 
\bibitem[Parker(1955)]{1955ApJ...122..293P} Parker, E.~N.\ 1955, \apj, 122, 293 

\bibitem[Parker(1993)]{1993ApJ...408..707P} Parker, E.~N.\ 1993, \apj, 408, 707

\bibitem[Prusa et al.(2008)]{prusa08} Prusa, J.~M., Smolarkiewicz, P.~K., Wyszogrodzki, A.~A.\ 2008, Comp. \& Fluids, 37, 1193 

\bibitem[Pipin \& Kosovichev(2011)]{2011ApJ...727L..45P} Pipin, V.V., and Kosovichev, A.~G.\ 2011, \apjl, 727, 45

\bibitem[Pipin \& Kosovichev(2019)]{2019arXiv190804525P} Pipin, V.V., and Kosovichev, A.~G.\ 2019, arXiv:1908.04525

\bibitem[Racine et al.(2011)]{Racine11} Racine, {\'E}., Charbonneau, P., Ghizaru, M., Bouchat, A., Smolarkiewicz, P.~K.\ 2011, \apj, 735, 46 

\bibitem[Schou et al.(1998)]{1998ApJ...505..390S} Schou, J., Antia, H.~M., Basu, S., et al.\ 1998, \apj, 505, 390

\bibitem[Smolarkiewicz \& Margolin(1998)]{smolar98} Smolarkiewicz, P.~K., Margolin, L.~G.\ 1998, J. Comp. Phys., 140, 459 

\bibitem[Smolarkiewicz \& Prusa(2002)]{smpr02} Smolarkiewicz, P.~K., Prusa, J.~M.\ 2002, {{in: D. Drikakis and B.J. Guerts (Eds.), Turbulent Flow Computation}}
Kluwer Academic Publishers, 279  

\bibitem[Smolarkiewicz(2006)]{smolar06} Smolarkiewicz, P.~K.\ 2006, IJNMF, 50, 1123 

\bibitem[Smolarkiewicz \& Charbonneau(2013)]{smolar13} Smolarkiewicz, P.~K., Charbonneau, P.\ 2013, J. Comp. Phys., 236, 608 

\bibitem[Steenbeck et al.(1966)]{1966ZNatA..21..369S} Steenbeck, M., Krause, F., R\"adler, K.~H.\ 1966, ZNatA, 21, 369

\bibitem[Strugarek et al.(2016)]{Strugarek16} Strugarek, A., Beaudoin, P., Brun, A.~S., et al.\ 2016, Advances in Space Research, 58, 1538 

\bibitem[Warnecke et al.(2013)]{2013ApJ...778..141W} Warnecke, J., K\"apyl\"a, P.~J., Mantere, M.~J., Brandenburg, A.\ 2013, \apj, 778, 141

\bibitem[Warnecke et al.(2018)]{2018A&A...609A..51W} Warnecke, J., Rheinhardt, M., Tuomisto, S., et al.\ 2018, \aap, 609, 51

\bibitem[Warnecke(2018)]{2018A&A...616A..72W} Warnecke, J.\ 2018, \aap, 616, 72 

\bibitem[Wray et al.(2015)]{2015arXiv150707999W} Wray, A.~A., Bensassi, K., Kitiashvili, I.~N., Mansour, N.~N., Kosovichev, A.~G.\ 2015, ArXiv:1507.07999

\bibitem[Zhao et al.(2013)]{2013ApJ...774L..29Z} Zhao, J., Bogart, R.~S., Kosovichev, A.~G., Duvall Jr., T.~L.,  Hartlep, T.\ 2013, \apjl, 774, 29

\end{thebibliography}
\end{document}